# Decision theory and information propagation in quantum physics

Alan Forrester

15 Linden Road

Northampton

NN3 2JJ

alan_forrester2@yahoo.co.uk

## Abstract

In recent papers, Zurek (2005) has objected to the decision-theoretic approach of (Deutsch, 1999) and (Wallace, 2003) to deriving the Born rule for quantum probabilities on the grounds that it courts circularity. Deutsch and Wallace assume that the many worlds theory is true and that decoherence gives rise to a preferred basis. However, decoherence arguments use the reduced density matrix, which relies upon the partial trace and hence upon the Born Rule for its validity. Using the Heisenberg Picture and quantum Darwinism – the notion that classical information is quantum information that can proliferate in the environment pioneered in (Olliver et al. 2005 and 2006) – I show that measurement interactions between two systems only create correlations between a specific set of commuting observables of system 1 and a specific set of commuting observables of system 2. This argument picks out a unique basis in which information flows in the correlations between those sets of commuting observables. I then derive the

Born rule for both pure and mixed states and answer some other criticisms of the decision theoretic approach to quantum probability.

## 1. Introduction

The basic equations of quantum mechanics are uncontroversial, but the explanation (interpretation) of these equations is still very controversial. According to the many worlds interpretation quantum physics should be taken literally – in particular when its equations say that physical objects exist in multiple versions this should be taken to mean that these other versions do actually exist (Everett, 1957; Deutsch, 1997 and 2002). One of the problems with this view is that one cannot experiment directly on these other versions of objects, but can only do so indirectly through interference. If the many worlds theory is true, that must presumably be because physical reality (in this context called the multiverse) is divided approximately into separate layers, each of which resembles the world of classical physics to some extent, for some demonstrable reason. One approach to securing this conclusion has been decoherence: every system interacts with an environment to some extent, and the hope was that by tracing the environmental degrees of freedom out of the density matrix of a system it could be shown that open quantum systems have a 'pointer basis' in which they act classically to a suitable degree of approximation. Calculations and computer simulations of decoherence seemed to indicate that there is a pointer basis for many interesting systems (Zurek, 1991). However, as Zurek (2005) pointed out, decoherence arguments assume that the partial trace procedure for extracting reduced density matrices is valid and this in turn assumes that the Born rule is valid (Neilsen and Chuang, 2000, p. 107; Landau, 1927). The Born Rule states that for a quantum observable $\hat{A}$ of a system in a state $\rho$ the expectation value of $\hat{A}$ is given by:

$$\langle \hat{A} \rangle = Tr(\rho \hat{A}) \tag{1}$$

where *Tr* is the trace function.

Another problem is that quantum systems do not obey the axioms of standard probability theory (see e.g. (Deutsch, Ekert and Luppachini, 2000) – for instance, there is no way of assigning probabilities obeying the chain rule, to values of observables at times when they are not 'observed'). Hence, the Born rule for obtaining probabilities from quantum systems is not generally applicable while interference processes are under way. So the question arises: under what circumstances do quantum systems obey the Born rule? There have been many unsuccessful approaches to this problem using the frequency interpretation of probability, which is unviable because it relies on the unphysical notion of an infinite set of measurements. More recently there have been some approaches to probability via decision theory (Deutsch, 1999; DeWitt, 1998; Wallace, 2003). Decision theory describes the behaviour of agents or decision makers who assign 'values' to different games in a way that satisfies the 'rationality' requirement, namely that for any two games $G_1$ and $G_2$ either the value of $G_1$ is greater than that of $G_2$, or it is less than that of $G_2$ or it is equal in value to $G_2$. If the value of $G_1$ is greater than that of $G_2$, the agent will willingly give up an opportunity to experience $G_2$ for an opportunity to experience $G_1$ and if the ranking is reversed his preferences will reverse too. If their value is equal he will be indifferent between experiencing $G_1$ and experiencing $G_2$. If, as in quantum physics, a game has many possible outcomes then he will assign probabilities to the different possible outcomes to assign the game a value. It may be thought that decision theory is irrelevant to the laws of physics. However this objection is wrong because any real agent must be a physical system. In particular, if the laws of physics allow the existence of agents who act in the manner described above, then decision theory is applicable to physics; otherwise decision theory is not applicable to physics. If decision theory is applicable to physics then in some sense the laws of rationality are laws of physics because they predict the behaviour of a certain physical systems – decision theoretic agents. This argument also answers the objection that the laws of physics, including the Born rule should underwrite decision theory and not the other way around. Previous quantum mechanical decision theoretic arguments assumed that decoherence selects a preferred basis that evolves classically to a good approximation and so that information could only propagate in this preferred basis. As agents have to be able to receive information and use it to change their environment – perhaps by placing bets – they could only care about a set of outcomes in the preferred basis. In this approach, it was argued that quantum physics in the Schrödinger Picture (with static observables and an evolving global state) is compatible with decision theory in the sense that under suitable circumstances the Born rule (and, indeed, only it) satisfies the rationality requirement and the axioms of probability theory in the Everett interpretation.

Zurek (2005) argues that the decision theoretic approach courts circularity because the decision-theoretic arguments proposed to date take for granted that decoherence provides a preferred 'pointer basis' and this relies on the partial trace which relies on the Born rule (Neilsen and Chuang, 2000, p.107; Landau, 1927). Zurek then derives the probability rule using either decision theory or the relative frequency approach, using certain invariance properties of Schrödinger states of systems that have been measured. He calls this approach *envariance*.

By contrast with Deutsch and Wallace, I will not assume that decoherence gives rise to a preferred basis. Instead I show that a measurement interaction between two systems only create correlations between observable $\hat{A}_1(t)$ of a system $S_1$ and observable $\hat{A}_2(t)$ of another system $S_2$. This picks out a unique basis in which information flows in the correlations between observables of these quantum systems. The properties of these measurement interactions place substantial restrictions on the information that can propagate in the environment and so on the information that an agent can get about a system and I use these restrictions to derive the Born rule by treating decision theoretic agents as information-processing systems. (Ollivier *et al.*, 2004 and 2005) have also discussed the constraints on what sort of information about a quantum system can spread widely in the environment. Unlike Ollivier *et al.*, I use the Heisenberg Picture because the Schrödinger picture (with its global but evolving state) is not as useful for the study of information propagation as is the Heisenberg Picture (with its static state and local evolving observables), as Deutsch and Hayden (2000) have pointed out. Of course, it is possible to trace the flow of information in the Schrödinger Picture as it is mathematically equivalent to the Heisenberg Picture, and it has been used to prove results about the flow of information in quantum systems in (Tipler, 2000; Page, 1982), so these results could be derived in the Schrödinger Picture, but the argument would be more difficult to follow. Ollivier *et al.* also discuss quantitative issues of how to decide how much information flows out of a system into the environment. In this paper I discuss only qualitative issues of quantum Darwinism that have a bearing on quantum decision theory. In Section 2 I derive the relevant part of quantum Darwinism using the Heisenberg Picture. In Section 3 I use the results of Section 2 together with decision theory to derive the Born rule. In Section 4 I discuss the result and some criticisms of the decision theoretic approach.

## 2. Quantum Darwinism in the Heisenberg picture

First I shall explain the formalism that I will use to obtain the results in this paper. In the Heisenberg Picture, an arbitrary observable $\hat{A}_1(t)$ of a system $S_1$ with an $N$-dimensional Hilbert space is a Hermitian operator which can be written as

$$\hat{A}_1(t) = \sum_{a=0}^{N-1} \alpha_{1a} \hat{B}_{1Aa}(t), \tag{2}$$

where the $\alpha_{1a}$ are static, real c-numbers and the $\hat{B}_{1Aa}(t)$ are Hermitian operators such that

$$\begin{aligned} \hat{B}_{1Aa}(t)\hat{B}_{1Ab}(t) &= \delta_{ab}\hat{B}_{1Aa}(t) \\ \sum_{a=0}^{N-1} \hat{B}_{1Aa}(t) &= \hat{1} \end{aligned}, \tag{3}$$

where $\hat{1}$ is the unit observable. Thus the $\hat{B}_{1Aa}(t)$ are a set of commuting projectors. (I have used this notation to avoid confusion between projectors and payoff functions, see Section 3.) I shall discuss he role of the static Heisenberg state below. Observables of $S_1$ that do not commute with $\hat{A}_1(t)$ have different sets of projectors from $\hat{A}_1(t)$, but the equation for those observables in terms of those different projectors has the same form of as equation (2).

So the $\hat{B}_{1Aa}(t)$ do not span the full vector space of the observables of $S_1$ so some more operators are needed. The operators $S_{1Aab}(t)$ are defined to have the properties

$$S_{1Aab}(t)S_{1Acd}(t) = \delta_{bc} S_{1Aad}(t)$$
$$S_{1Aaa}(t) = \hat{B}_{1Aa}(t)$$
(4)

In other words, these operators have the same algebra as matrices with a 1 in the $(a,b)$th slot and 0 elsewhere. The $S_{1Aab}(t)$ are a full basis of the vector space of the observables of $S_1$. (However, the $S_{1Aab}(t)$ are not all observables since they are not all Hermitian.) An arbitrary observable $\hat{C}_1(t)$ of $S_1$ that is not equal to $\hat{A}_1(t)$ may be written as

$$\hat{C}_1(t) = \sum_{c=0}^{N-1} \gamma_{1c} \hat{B}_{1Cc}(t)$$
$$\hat{B}_{1Cc}(t) = \sum_{c=0}^{N-1} \beta_{1cde} S_{1Ade}(t)$$
(5)

where the $\gamma_{1c}$ are real, static c-numbers (the eigenvalues of $\hat{C}_1(t)$) and the $\beta_{1cde}$ are complex static c-numbers such that the projectors $\hat{B}_{1Cc}(t)$ obey the algebra (3).

An arbitrary observable $\hat{A}_1(t)$ evolves as

$$\hat{A}_1(t) = U_t^\dagger \hat{A}_1(0) U_t$$
$$U_t^\dagger U_t = U_t U_t^\dagger = \hat{1}$$
(6)

where $U_t^\dagger$ is the Hermitian conjugate of $U_t$. $U_t$ may or may not depend only on descriptors of $S_1$. The $S_{1Aab}(t)$ are linear combinations of observables and so they also evolve according to this rule. I will suppose for the rest of this paper, with no relevant loss of generality, that the systems investigated evolve in steps of unit duration and the observables will be considered only at the end of each step. The generality follows from the fact that a quantum computational network evolving in the manner described may simulate any transformation undergone by any finite quantum system with arbitrary accuracy (Deutsch, 1989). Nothing would be lost by discarding this assumption, except for some clarity. A computer is a physical system composed of subsystems $S_1, S_2...$ with associated observable quantities that can take on some discrete range of values. The computer evolves in discrete steps during which only some finite number of subsystems interact with one another. The motion undergone by the computer is called a computation.

I shall now explain in what sense a quantum system can perform a classical computation – that is, a computation that could be performed by a Turing Machine obeying classical physics as defined by Turing (1936) – and how information propagates between quantum systems. Consider a specific example of motion of a quantum system:

$$U = \sum_{b=0}^{N-1} \exp(i\phi_b) S_{1 A b \, \pi(b)}(0), \tag{7}$$

where $\pi$ is a permutation of the integers $0...N-1$ and the $\phi_b$ are arbitrary real numbers. Then

$$\hat{A}_1(1) = \sum_{a=0}^{N-1} \alpha_{1a} \hat{B}_{1 A \, \pi(a)}(0), \tag{8}$$

so this motion has just permuted the eigenvalues corresponding to each of the projectors associated with $\hat{A}_1(t)$ and the phases $\phi_b$ don't matter so far as the evolution of this observable under $U$ is concerned.

However, $U$ has a very different effect on the $\hat{B}_{1Cc}(t)$ (and hence on the arbitary observable $\hat{C}_1(t)$ provided it does not commute with $\hat{A}_1(t)$):

$$\hat{B}_{1Cc}(1) = \sum_{c=0}^{N-1} \exp[i(\phi_d - \phi_e)] \beta_{1cde} S_{1A\,\pi(d)\,\pi(e)}(0). \tag{9}$$

The phases $\phi_b$ do matter to $\hat{C}_1(t)$ and this evolution does not permute the $\hat{B}_{1Cc}(t)$.

Since the evolution of $\hat{A}_1(t)$ under $U$ values permutes projectors that can be labelled with the integers from 0 to $N-1$ and any algorithm that permutes integers from 0 to $N-1$ could be performed by a Turing machine obeying classical physics, the evolution of $\hat{A}_1(t)$ under $U$ performs $N$ classical computations evaluating the same function of the integers from 0 to $N-1$ with different initial values. Each of these is locally and possibly globally a branch of the multiverse – that is to say, a structure within which information flows perfectly about values of observables $\hat{A}_1(t)$ at an earlier time. (See (Deutsch, 2002) for an elaboration of this argument.) A branch of the multiverse must be a structure within which information flows because one of the characteristic features of the macroscopic world is that information does flow in the sense that it is possible to infer some features of the past from records and to predict some features of the future from the present state of the world. Furthermore as I show below information flows perfectly between systems only in branches characterised by classical computation. The evolution of $\hat{C}_1(t)$ under $U$ does not perform a classical computation and does not constitute a branch of the multiverse. As I will discuss below a system may branch in different ways at different times so a branch may not survive for very long. If a branch continues to exist on large 'classical' scales I will call it a 'world'.

In the many worlds theory, a measurement of a quantum system produces correlations between observables of the measured system and the measuring system and does not select one possible value of that observable as "the value" of that observable. Nevertheless, observers and measuring instruments do not observe all of the possible values of an observable.

The standard proof that, in the many worlds interpretation, observers and measuring instruments do not observe all of the possible values of an observable runs as follows. Let $S_1$ and $S_2$ be two systems with $N$-dimensional Hilbert spaces $H_1$ and $H_2$ respectively with the joint system $S_1 S_2$ having Hilbert space $H_1 \otimes H_2$. Consider the observables $\hat{A}_1(t)$ and $\hat{A}_2(t)$ of $S_1$ and $S_2$ respectively

$$\hat{A}_1(t) = \sum_{a=0}^{N-1} \alpha_{1a} \hat{B}_{1Aa}(t)$$
$$\hat{A}_2(t) = \sum_{b=0}^{N-1} \alpha_{2b} \hat{B}_{2Ab}(t)$$
(10)

where these two observables have identical spectra. And let

$$\hat{C}_1(t) = \sum_{c=0}^{N-1} \gamma_{1c} \hat{B}_{1Cc}(t)$$

$$\hat{B}_{1Cc}(t) = \sum_{d,e=0}^{N-1} \beta_{1cde} S_{1Ade}(t)$$

$$\hat{C}_2(t) = \sum_{c=0}^{N-1} \gamma_{2c} \hat{B}_{2Cc}(t) \quad , \tag{11}$$

$$\hat{B}_{2Cc}(t) = \sum_{d,e=0}^{N-1} \beta_{1cde} S_{1Ade}(t)$$

where the observables $\hat{C}_1(t)$ and $\hat{C}_2(t)$ have the same spectra as the observables $\hat{A}_1(t)$ and $\hat{A}_2(t)$ but do not commute with $\hat{A}_1(t)$ and $\hat{A}_2(t)$.

Consider the motion described by

$$U_{M1} = \sum_{a,b=0}^{N-1} \exp(i\phi_{ab}) \hat{B}_{1Aa}(0) S_{2Aba \oplus_N b}(0), \tag{12}$$

where

$$a \oplus_N b \equiv (a+b) \bmod N . \tag{13}$$

Then

$$\hat{A}_1(1) = \sum_{a=0}^{N-1} \alpha_{1a} \hat{B}_{1Aa}(0) = \hat{A}_1(0)$$

$$\hat{A}_2(1) = \sum_{a,b=0}^{N-1} \alpha_{2b} \hat{B}_{1Aa}(0) \hat{B}_{2Aa \oplus_N b}(0) \tag{14}$$

So $\hat{A}_1(t)$ is unaffected but $\hat{A}_2(t)$ now depends on the values of $a$ and $b$ and so this is called a perfect measurement of $\hat{A}_1(t)$ by $\hat{A}_2(t)$ as it perfectly correlates $\hat{A}_2(t)$ with $\hat{A}_1(t)$. Furthermore, (12) performs a classical computation on $\hat{A}_1(t)$ and $\hat{A}_2(t)$ in the sense described above. As such there are $N$ independent branches and in each version the measuring instrument measures only one version of the measured system.

However, $\hat{C}_1(t)$ and $\hat{C}_2(t)$ respond very differently:

$$\hat{C}_1(1) = \sum_{a,b,c,d=0}^{N-1} \exp[i(\phi_{db} - \phi_{ab})] \gamma_{1c} \beta_{1cad} S_{1Aad}(0) S_{2Aa\oplus_N bd\oplus_N b}(0)$$
$$\hat{C}_2(1) = \sum_{a,b,c,d=0}^{N-1} \exp[i(\phi_{ad} - \phi_{ab})] \gamma_{2c} \beta_{2cbd} \hat{B}_{1Aa}(0) S_{2Aa\oplus_N ba\oplus_N d}(0)$$

(15)

$\hat{C}_1(t)$ and $\hat{C}_2(t)$ are not perfectly correlated with one another or with anything else and do not evolve classically under (12). Thus (12) perfectly correlates $\hat{A}_1(t)$ and $\hat{A}_2(t)$, but not $\hat{C}_1(t)$ and $\hat{C}_2(t)$. Hence the form of the measurement interaction selects a preferred pointer basis: it induces the existence of branches down which information flows characterised by different values of $\hat{A}_1(t)$, but not of $\hat{C}_1(t)$. Furthermore, (14) and (15) illustrate that measurement can only copy information about eigenvalues of commuting sets of observables. Since information does get copied from one system to another (e.g. – from the document I am writing to other identical copies of this document that other people will read) and measurements of the kind described above are classical computations on $\hat{A}_1(t)$ and $\hat{A}_2(t)$ branches must be characterised by classical evolution.

What happens if $\hat{C}_1(t)$, which isn't correlated with anything is measured onto $\hat{A}_2(t)$? First, between times 1 and 2, $S_1$ evolves so that

$$\hat{A}_1(2) = \hat{C}_1(1).$$

(16)

If the measurement

$$U_{M2} = \sum_{a,b=0}^{N-1} \exp(i\phi_{ab}) \hat{B}_{a1}(2) S_{2Aba\oplus_N b}(2) \qquad (17)$$

is then performed the result is

$$\hat{A}_2(3) = \sum_{a,b=0}^{N-1} \alpha_{2b} \hat{B}_{1Aa}(2) \hat{B}_{2Aa\oplus_N b}(2)$$

$$= \sum_{a,b,c,d,e=0}^{N-1} \alpha_{2b} \beta_{1a\,e\,a\oplus_N b} \exp[i(\phi_{a\oplus_N b\,d} - \phi_{cd})] S_{1A\,e\,a\oplus_N b}(0) S_{2A\,e\oplus_N d\,a\oplus_N b\oplus_N c}(0) \qquad (18)$$

This is effectively a measurement of $\hat{C}_1(t)$ onto $\hat{A}_2(t)$. From (18) it can be seen that the observable still has a branching structure in a suitably chosen basis but each branch that existed at time 0 is not associated with a single branch at time 2 but instead with multiple branches. As such not all the classical information that was recorded in $\hat{A}_2(t)$ before the measurement has been retained.

So if measurements are always made in the same basis then the measured observable and the measuring observable may be regarded as a perfect channel for information about the results of classical computations. Otherwise it may not. And that is the main point of quantum Darwinism. Even here, however, the observables will carry $N$ versions of this information as all classical evolution changes all $N$ versions of a particular system. Quantum Darwinism in the sense used here is somewhat stricter than the sense in which the term is used in (Ollivier *et al.*, 2004 and 2005). I discuss the issue of the relationship between the two ideas in Section 4.

A system that processes classical information – such as the human brain, or a decision theoretic agent – only does so in some of its observables. Thus, a classical computation, such as human thought, must be instantiated only on some of the observables describing the physical object performing the computation. Even a quantum computer can only contain an answer to a problem in some commuting set of observables if that answer is to be propagated and used. And in general since the classically evolving observable is doing many classical computations an observer can experience only one of these computations, the particular computation that instantiates him, and so each copy of the observer can observe only one of the possible values of an observable after a measurement. The constant Heisenberg relative state $\rho$, a positive operator with trace 1, contains information about what observables carry information and what their values are in the world under consideration. This information is necessary for making predictions. In this sense, the state is an emergent feature of information flow between observables in quantum systems. The existence of the state is a consequence of measurement theory in the many worlds theory and is not needed to derive measurement theory. So quantum Darwinism is more fundamental than envariance, which relies on the existence of the relative state. The information that can be accessed in the state $\rho$ by an observer about an observable $\hat{A}_1(t)$ is contained in $\rho\hat{A}_1(t)$. And since measurement has the form of classical evolution in the sense described above an observer cannot access any phase information in $\rho\hat{A}_1(t)$. When some observable has a definite value in the world under consideration the state is said to be pure. A pure state obeys the equation $\rho^2 = \rho$ and so the state is equal to the product of projectors for some set of observables at equal times for all of the subsystems of the system of interest. For a pure state there is an observable $\hat{A}_1(t)$ of the whole system for which

$$\rho\hat{A}_1(t) = \alpha_{1a}\rho, \tag{19}$$

In all of the worlds in the region of the multiverse described by the state $\rho$ the observable $\hat{A}_1(t)$ has value $\alpha_{1a}$. If there is no such observable then the state is said to be mixed.

I will need one more form of measurement to provide a proof of the Born rule. Sometimes an observable on an $N$-dimensional Hilbert space $H_1$ may be measured onto an observable of an $M$-dimensional Hilbert space $H_1$ with $M > N$. To do this $H_2$ is divided into $a$ subspaces $H_{2a}$ of dimension $m_a$ so that

$$H_2 = \otimes_{a=0}^{N-1} H_{2a}. \tag{20}$$

The observables $\hat{A}_1(t)$ and $\hat{A}_2(t)$ with spectra $\alpha_{1a}$ and $\alpha_{2b}$ with $a = 0...N-1, b = 0...M-1$ respectively then evolve such that

$$\hat{A}_1(t+1) = \sum_{a=0}^{N-1} \alpha_{1a} \hat{B}_{1Aa}(t)$$

$$\hat{A}_2(t+1) = \sum_{a=0}^{N-1} \hat{B}_{1Aa}(t) \left( \sum_{b=\gamma_{a-1}}^{\gamma_a} \alpha_{2b} \hat{B}_{2Ab}(t) \right). \tag{21}$$

$$\gamma_j = \sum_{k=0}^{j} m_k$$

This measurement differs from the ones above only in that for every possible value $\alpha_{1a}$ many possible values of the measuring observable act as indicators of that value of $\hat{A}_1(t)$. This kind of measurement gives the same information as the earlier simpler measurements like (17) and (12).

## 3. Decision theory and quantum probability

Decision theory concerns rational agents, that is, physical objects such as persons that rank different physical states of affairs on a single scale of value. These agents may be considered as playing games in which each possible consequence of a particular action has some payoff associated with it. The value of a game is the payoff such that the agent is neutral between receiving that payoff and playing the game. In the context of quantum theory an agent is an information-processing system that obeys quantum physics and a physical state of affairs consists of an observable to be measured and a state in which it will be measured. By using this approach do I presuppose the Born rule? Unitary evolution does not presuppose the Born rule and indeed some unitary evolution changes amplitudes in ways that violate the Born rule as mentioned in Section 1. The measurement theory of Section 2 is about the creation of perfect correlations between observables and so explains what information an agent can extract from a system. In Section 2 I made no reference to probability and did not use the Born rule.

More explicitly, a quantum game consists of (1) an observable to be measured $\hat{A}_1(t)$, the measurement results being the consequences of the agent's decision to play that particular game, (2) a state $\rho$ in which $\hat{A}_1(t)$ is measured, (3) a payoff function $P$ that gives information on how much the agent gets for each eigenvalue of $\hat{A}_1(t)$ that is observed (i.e. – it is a function from the eigenvalues of $\hat{A}_1(t)$ to the real numbers) and (4) a value function $V$ which maps $\hat{A}_1(t)$, the state $\rho$ and the payoff function $P$ to a real number that is the value of the game.

In addition there is a set of acts that may be performed upon the system $S_1$ that has observable $\hat{A}_1(t)$ and ancillary systems to change one game into another – unitary operations.

I should briefly discuss assumption (3) because some physicists apparently regard it as problematic. For an example see (Barnum et al., 2000, Section III) in which they complain that Deutsch assumes that when the eigenvalues are the same in each branch the player always gets the same payoff. As noted above when one observable is measured onto another, the measurement only conveys information about that observable and those observables of the measured system with which it commutes. Moreover, in each branch only information about the specific value that has been measured in that branch is available and can affect what happens in that branch. So attempts to argue that payoffs should not be associated with eigenvalues of the measured observable are perverse.

The result that I will derive is

$$V(\rho, \hat{A}_1(t), P_0) = Tr(\rho \hat{A}_1(t)), \tag{22}$$

where $P_0$ is the payoff function that gives payoffs equal to the eigenvalues of $\hat{A}_1(t)$ in each branch. That is, the expectation value $Tr(\rho \hat{A}_1(t))$ predicted by the Born Rule will be shown to be the same as the value of the game in state $\rho$, observable $\hat{A}_1(t)$ and payoff function $P_0$. So any quantum measurement may be interpreted as a quantum game with specific payoffs and expectation values.

If a particular game is changed in any of the attributes listed (1) – (3) then it becomes a different game. If two games have the same value they may be said to be equivalent because the agent is indifferent between playing them. The following properties of the value function follow from the general description of a game above and the discussion of Section 2:

An agent must be able to retain information about the value of an observable he has experienced if he is to keep the payoff associated with that value. Otherwise he will forget the value and the associated payoff. So he must get the payoff by measuring commuting observables and keeping the records in observables that commute with the observables that will be measured by other systems. And as each memory record within the agent may be measured by other observables that constitute the agent, all the observables in which the records are kept must commute with one another. This means that the agent cannot care about phases in the product $\rho \hat{A}_1(t)$ that provides him with all the information he can access about $\hat{A}_1(t)$. This property of rational agents is called **phase indifference**.

If an act $U$ is a classical computation relative to the observable $\hat{A}_1(t)$ to be measured and $U^\dagger \rho U = U \rho U^\dagger = \rho$ then $U$ does not change the game's value and I will call $U$ a classical act. $U$ just permutes which of the possible values of the observable are associated with which branch with respect to the observable concerned and with respect to the state picking out the world in which the game takes place. That is, the agent cares about the consequences associated with being in a branch, not how the branches happen to be labelled. This principle is called **classical act neutrality**.

**Physicality** If there are two observables $\hat{A}_1(t)$ and $\hat{A}_2(t)$ such that

$$\rho \hat{A}_1(t+\Delta t) = \rho \hat{A}_2(t), \tag{23}$$

then

$$V\left(\rho, \hat{A}_1(t+\Delta t), P\right) = V\left(\rho, \hat{A}_2(t), P\right), \tag{24}$$

where $P$ is an arbitrary payoff function.

I have argued that all of the information available about an arbitrary observable $\hat{A}_1(t)$ is contained in $\rho \hat{A}_1(t)$ and the physicality principle is a direct consequence of this.

**Dominance** Let $P_1$ and $P_2$ be payoff functions such that $P_1 \geq P_2$ for all members of the spectrum of $\hat{A}_1(t)$ then

$$V\left(\rho, \hat{A}_1(t), P_1\right) \geq V\left(\rho, \hat{A}_1(t), P_2\right). \tag{25}$$

This amounts to the reasonable assumption that if in every world a player gets a greater payoff when he plays game $G_1$ then he does if he plays game $G_2$ then he will prefer to play game $G_1$.

**Additivity** Let $P_1$ and $P_2$ be arbitrary payoff functions, then

$$V\left(\rho, \hat{A}_1(t), P_1 + P_2\right) = V\left(\rho, \hat{A}_1(t), P_1\right) + V\left(\rho, \hat{A}_1(t), P_2\right). \tag{26}$$

This means that placing one bet whose payoff function is $P_1 + P_2$ on a particular physical situation (i.e. – a particular instance of $\rho$ and $\hat{A}_1(t)$) is the same as placing two separate bets on the same physical situation whose payoff functions are $P_1$ and $P_2$ respectively. To see why this is true consider a game with state $\rho$, observable $\hat{A}_1(t)$ and payoff function $P_1$. This is equivalent to measuring the observable $P_1(\hat{A}_1(t))$

$$P_1(\hat{A}_1(t)) = \sum_{a=0}^{N-1} P_1(\alpha_{1a}) \hat{B}_{1Aa}(t) \tag{27}$$

with the payoff function $P_0$ since these give the same payoff in every branch. Similarly for another payoff function $P_2$ measuring $P_2(\hat{A}_1(t))$ and getting the payoff $P_0$ is the same as measuring $\hat{A}_1(t)$ and getting payoff $P_2$. The observables $\hat{A}_1(t)$, $P_1(\hat{A}_1(t))$, $P_2(\hat{A}_1(t))$ and $(P_1 + P_2)(\hat{A}_1(t))$ all commute with one another and so records of the results of measuring all of these observables do not conflict with one another. So an agent could measure $P_1(\hat{A}_1(t))$ and $P_2(\hat{A}_1(t))$ separately with payoff $P_0$ and receive a payoff $(P_1 + P_2)(\alpha_{1a})$ in the $a$th branch just as he would if he measured $(P_1 + P_2)(\hat{A}_1(t))$.

I will now prove the Born rule in four stages, the first three of which concern only the value of games played with pure states. All of the games below will consist of a measurement of $\hat{A}_1(0)$ in the state $\rho$ with payoff function $P_0$ unless otherwise specified.

**Stage 1 Equal probabilities** I now follow (Wallace, 2003) to prove that decision theory agrees with the Born Rule in cases where all results occur with equal probability. To do this I consider the game with

$$\rho = \frac{1}{N} \sum_{a,b=0}^{N-1} S_{1A\,ab}(0)$$
$$\hat{A}_1(0) = \sum_{a=0}^{N-1} \alpha_{1a} \hat{B}_{1A\,a}(0)$$
(28)

From (28)

$$\rho \hat{A}_1(0) = \frac{1}{N} \sum_{a,b=0}^{N-1} \alpha_{1b} S_{1Aab}(0).$$
(29)

I will now show that the value of this game is the same as the value of another game, which has the same value in all branches of the multiverse – the value given by the Born rule. Define $P_\pi$ as the payoff function such that $P_\pi(\alpha_{1a}) = \alpha_{1\pi(a)}$ where $\pi$ is a permutation of the integers $0...N-1$. Let $U_{\pi^{-1}}$ be the unitary transformation that performs the permutation $\pi^{-1}$ on the projectors of $\hat{A}_1(t)$ then

$$\begin{aligned}V(\rho, \hat{A}_1(t), P_\pi) &= V\left(\rho, \sum_{a=0}^{N-1} \alpha_{1\pi(a)} \hat{B}_{1A\,a}(0), P_0\right) \\ &= V\left(\rho, \sum_{a=0}^{N-1} \alpha_{1a} \hat{B}_{1A\,\pi^{-1}(a)}(0), P_0\right) \\ &= V\left(\rho, U_{\pi^{-1}} \hat{A}_1(t) U_{\pi^{-1}}^\dagger, P_0\right) \\ &= V(\rho, \hat{A}_1(t), P_0)\end{aligned}$$
(30)

The first equality follows from the definitions of $P_\pi$ and $P_0$. The second and third are trivial. The fourth is a consequence of classical act neutrality.

Consider the game that has the payoff function $\sum_\pi P_\pi$. From (30) the value of this game is

$$V\left(\rho, \hat{A}_1(t), \sum_\pi P_\pi\right) = N! V\left(\rho, \hat{A}_1(t), P_0\right) \tag{31}$$

since there are $N$ permutations of $0...N-1$. The function $\sum_\pi P_\pi$ assigns the same payoff to every branch, that is

$$V\left(\rho, \hat{A}_1(t), \sum_\pi P_\pi\right) = (N-1)! \sum_{a=0}^{N-1} \alpha_{1a}. \tag{32}$$

To see this, consider the $j$th branch. How often does $\alpha_{1j}$ appear in this branch? As many times as there are permutations that have the effect $\alpha_{1j} \to \alpha_{1j}$; that is to say it appears $(N-1)!$ times. (31) and (32) give

$$V\left(\rho, \hat{A}_1(t), P_0\right) = \frac{1}{N} \sum_{a=0}^{N-1} \alpha_{1a} \tag{33}$$

This is the result expected from the Born rule.

**Stage 2 Unequal but rational probabilities** I will now show that in cases where the probabilities are unequal but rational, decision theory predicts the same expectation value as the Born Rule. Consider the game

$$\begin{aligned}
\rho &= \frac{1}{M} \sum_{a,b=0}^{N-1} \sqrt{m_a m_b} S_{1Aab}(0) \sum_{c,d=0}^{M-1} \frac{1}{M} S_{2Acd}(0) \\
\hat{A}_1(t) &= \sum_{a=0}^{N-1} \alpha_{1a} \hat{B}_{1Aa}(t) \\
\hat{A}_2(t) &= \sum_{e=0}^{M-1} \alpha_{2e} \hat{B}_{2Ae}(t) \\
\sum_{a=0}^{M-1} m_a &= M
\end{aligned} \tag{34}$$

Suppose, further, that the first $m_1$ eigenvalues of $\hat{A}_2(t)$ are $\alpha_{11}$, that the next $m_2$ eigenvalues of $\hat{A}_2(t)$ are $\alpha_{12}$ and so on.

I will now show that the value of this game is the same as the value of another game, a game in which the probabilities are all equal but which has the value given by the Born rule for the unequal but rational probability game. A measurement of the form (21) is performed from $\hat{A}_1(t)$ to $\hat{A}_2(t)$ at time 0, so that

$$\hat{A}_1(1) = \sum_{a=0}^{N-1} \alpha_{1a} \hat{B}_{1Aa}(0) = \hat{A}_1(0)$$

$$\hat{A}_2(1) = \sum_{a=0}^{N-1} \hat{B}_{1Aa}(0) \sum_{e=\gamma_{a-1}}^{\gamma_a} \alpha_{2e} \hat{B}_{1Ae}(0). \tag{35}$$

$$\gamma_j = \sum_{k=0}^{j} m_k$$

From (34) and (35)

$$\rho \hat{A}_1(1) = \rho \hat{A}_1(0) = \frac{1}{M} \sum_{a,b=0}^{N-1} \sqrt{m_a m_b} \, \alpha_{1b} S_{1Aab}(0) \sum_{c,d=0}^{M-1} \frac{1}{M} S_{2Acd}(0)$$

$$\rho \hat{A}_2(1) = \frac{1}{M} \sum_{a,b=0}^{N-1} \sqrt{m_a m_b} \, S_{1Aab}(0) \sum_{c=0}^{M-1} \sum_{e=\gamma_{a-1}}^{\gamma_a} \frac{1}{M} \alpha_{2e} S_{2Ace}(0) \tag{36}$$

And from the properties of the spectrum of $\hat{A}_2(t)$ the second equation of (36) may be rewritten as

$$\rho \hat{A}_2(1) = \frac{1}{M} \sum_{a,b=0}^{N-1} \sqrt{m_a m_b} \, \alpha_{1b} S_{1Aab}(0) \sum_{c,d=0}^{M-1} \frac{1}{M} S_{2Acd}(0) = \rho \hat{A}_1(0). \tag{37}$$

It follows from (37) and Physicality that

$$V(\rho, \hat{A}_2(1), P_0) = V(\rho, \hat{A}_1(0), P_0). \tag{38}$$

(36) may be rewritten as

$$\rho \hat{A}_2(1) = \hat{B}_{1Cc}(0) \sum_{c,e=0}^{M-1} \frac{1}{M} \alpha_{2e} S_{2Ace}(0), \tag{39}$$

where $\hat{B}_{1Cc}(0)$ is a projector of some observable on $S_1$. And the invariance of $S_{1Aab}(0)$ operators on $H_1$ under permutations of $S_{2Aab}(0)$ operators, along with (33), (38), (39) and the properties of the spectrum of $\hat{A}_2(t)$ gives

$$V(\rho, \hat{A}_1(0), P_0) = \sum_{a=0}^{N-1} \frac{m_a \alpha_{1a}}{M}. \tag{40}$$

This is the result expected from the Born rule.

**Stage 3 Irrational probabilities** The value of a game with irrational probabilities follows from the result for rational probabilities plus dominance and the denseness of the rational numbers in the reals. Explicitly, following (Deutsch, 1999), suppose that $\hat{A}_1(t)$ at time 0 is such that the state $\rho$ sets the probability of some of its eigenvalues to 0 and some or all of the probabilities of the eigenvalues are irrational. Suppose that $U_{upper}$ is an evolution that changes $\hat{A}_1(t)$ so that an eigenvalue greater than all of the ones that have non-zero probability at time 0 has anon-zero rational probability. The value of this game is greater than that of the game with $\hat{A}_1(t)$ and $\rho$ at time zero. The value of the game at time 0 sets a lower bound on the value of all games of this form.

Similarly there are transformations $U_{lower}$ that change $\hat{A}_1(t)$ so that an eigenvalue less than all of the ones that have zero probability at time 0 has a non-zero rational probability. The value of the game at time 0 sets an upper bound on the value of all games of this form.

The values of the $U_{upper}$ games are given by the Born rule, as are those of the $U_{lower}$ games. So the Born rule also gives the value of games with irrational probabilities.

**Stage 4 Mixed states** Mixed states usually arise in the context of an open quantum system that is part of a larger system in a pure state. In this case, the probabilistic predictions concerning the open subsystem may be derived by using the above argument on the larger system of which the open subsystem is a part. But some authors, such as Deutsch (1991) and Hawking (1976), have postulated mixed states for the multiverse as a whole, so I shall give a direct proof that the Born Rule also works for mixed states:

**Stage 4.1 Mixed states and unsharp observables** Any mixed state is represented by an operator that is mathematically the same as a sum of some set of pure states with real coefficients that sum to 1. Suppose that an observable $\hat{A}_1(t)$ is measured in a mixed state and that the observable is unsharp with respect to each of the pure states in the sum representing the mixed state, then the density operator may be expressed in the same form as an observable that does not commute with $\hat{A}_1(t)$ and whose eigenvalues sum to 1:

$$\rho = \sum_{b=1}^{N} \mu_b \hat{B}_{1Cb}(0) = \sum_{b=1}^{N} \mu_b \beta_{1Cde} S_{1Ade}(0)$$
$$\hat{A}_1(0) = \sum_{a=1}^{N} \alpha_{1a} \hat{B}_{1Aa}(0) \qquad (41)$$
$$\sum_{c=1}^{N} \mu_c = 1$$

From (41)

$$\rho \hat{A}_1(0) = \sum_{b=0}^{N-1} \mu_b \beta_{1bde} \alpha_{1e} S_{1Ade}(0). \tag{42}$$

From phase indifference and physicality this is the same as a game with any state $\rho'$ such that

$$\rho' \hat{A}_1(0) = \sum_{d,e=1}^{N} \lambda_{de} \alpha_{1e} S_{1Ade}(0)$$

$$\lambda_{de} = \left| \sum_{b=0}^{N-1} \mu_b \beta_{1bde} \right| < 1 \tag{43}$$

From physicality, the value of this game is the same as the value of a game played with a suitable pure state and an observable that is unsharp relative to that pure state, which is given by the Born Rule as shown in Stages 1-3. Sometimes the state has the same projectors as the measured observable and a different argument is needed to cover these cases.

**Stage 4.2 Equal probabilities** I will now show that if an observable $\hat{A}_1(t)$ is measured in a mixed state and that the observable is sharp with respect to each of the pure states in the sum representing the mixed state and that the Born rule states that the probabilities of each outcome should be equal then the expected value of the resulting game is given by the Born rule. In this game, the observable and state are:

$$\rho = \frac{1}{N} \sum_{a=0}^{N-1} \hat{B}_{1Aa}(0)$$

$$\hat{A}_1(t) = \sum_{a=0}^{N-1} \alpha_{1a} \hat{B}_{1Aa}(0) \tag{44}$$

So all of the information accessible to an agent playing the game is contained in the expression:

$$\rho\hat{A}_1(t) = \frac{1}{N}\sum_{a=0}^{N-1}\alpha_{1a}\hat{B}_{1Aa}(0). \tag{45}$$

The argument given in Stage 1 above obviously works just as well on (45) to give

$$V(\rho,\hat{A}_1(t),P_0) = \frac{1}{N}\sum_{a=0}^{N-1}\alpha_{1a}, \tag{46}$$

which is the result expected from the Born Rule.

**Stage 4.3 Unequal rational probabilities** I will now show that if an observable $\hat{A}_1(t)$ is measured in a mixed state and that the observable is sharp with respect to each of the pure states in the sum representing the mixed state and that the Born rule states that the probabilities of each outcome should be equal then the expected value of the resulting game is given by the Born rule. To derive the value of this game I use the systems with the state and observables:

$$\begin{aligned}
\rho &= \sum_{a,b=0}^{N-1}\frac{m_a}{M}\hat{B}_{1Aa}(0)\sum_{b=0}^{M-1}\frac{1}{M}\hat{B}_{2Ab}(0) \\
\hat{A}_1(t) &= \sum_{a=0}^{N-1}\alpha_{1a}\hat{B}_{1Aa}(t) \\
\hat{A}_2(t) &= \sum_{b=0}^{M-1}\alpha_{2b}\hat{B}_{2Ab}(t) \\
\sum_{a=0}^{M-1}m_a &= M
\end{aligned} \tag{47}$$

Suppose, further, that the first $m_1$ eigenvalues of $\hat{A}_2(t)$ are $\alpha_{11}$, that the next $m_2$ eigenvalues of $\hat{A}_2(t)$ are $\alpha_{12}$ and so on.

As in Stage 2 I will show that the value of this game is the same as the value of another game, a game in which the probabilities are all equal but which has the value given by the Born rule for the unequal but rational probability game. From (47)

$$\rho \hat{A}_1(t) = \sum_{a,b=0}^{N-1} \alpha_{1a} \frac{m_a}{M} \hat{B}_{1Aa}(0) \sum_{b=0}^{M-1} \frac{1}{M} \hat{B}_{2Ab}(0)$$

$$\rho \hat{A}_2(t) = \sum_{a,b=0}^{N-1} \frac{m_a}{M} \hat{B}_{1Aa}(0) \sum_{b=0}^{M-1} \alpha_{2b} \frac{1}{M} \hat{B}_{2Ab}(0)$$

(48)

A measurement of the form (21) is performed from $\hat{A}_1(t)$ to $\hat{A}_2(t)$ at time 0, so that

$$\ddot{\mathcal{A}}_1(1) = \sum_{a=0}^{N-1} \alpha_{1a} \ddot{\mathcal{B}}_{1Aa}(0) = \ddot{\mathcal{A}}_1(0)$$

$$\ddot{\mathcal{A}}_2(1) = \sum_{a=0}^{N-1} \ddot{\mathcal{B}}_{1Aa}(0) \sum_{e=\gamma_{a-1}}^{\gamma_a} \alpha_{2e} \ddot{\mathcal{B}}_{2Ae}(0).$$

(49)

$$\gamma_j = \sum_{k=0}^{j} m_k$$

(49), (47) and the properties of the spectrum of $\hat{A}_2(t)$ give:

$$\rho \hat{A}_2(1) = \sum_{a,b=0}^{N-1} \frac{m_a}{M} \hat{B}_{1Aa}(0) \sum_{c,d=0}^{M-1} \frac{1}{M} \alpha_{2b} \hat{B}_{2Ab}(0)$$

$$= \sum_{a,b=0}^{N-1} \frac{m_a}{M} \alpha_{1a} \hat{B}_{1Aa}(0) \sum_{c,d=0}^{M-1} \frac{1}{M} \hat{B}_{2Ab}(0) = \rho \hat{A}_1(0)$$

(50)

From (50)

$$\rho \hat{A}_1(0) = \rho_1 \sum_{c,d=0}^{M-1} \frac{1}{M} \alpha_{2b} \hat{B}_{2Ab}(0),$$

(51)

where $\rho_1$ is a density operator on $H_1$. So permutations of the $\hat{B}_{2Ab}(0)$ do not affect $\rho_1$ and using physicality the value of the game in (51) is the same as the value of the game played by measuring $\hat{A}_2(t)$ in the state $\rho$. This argument gives the result:

$$V(\rho, \hat{A}_1(t), P_0) = \sum_{a=0}^{N-1} \frac{m_a}{M} \alpha_{1a}.$$

(52)

**Stage 4.3 Irrational probabilities** The argument that gives Born rule for mixed states with irrational probabilities follows exactly the same pattern as Stage 3 using the value of the games 4.1 - 4.3.

4. Discussion and conclusions

I have shown that under the many worlds theory, the Born rule correctly predicts the decision theoretic probabilities of the outcome of a measurement. I shall now discuss some criticisms of the decision theoretic approach to probability in the many world theory. Lewis (2005) and Greaves (2004) have pointed out that the many worlds theory does not allow the agent to be uncertain about the future given that he knows the initial conditions and the equation of motion. Wallace (2006) objects that this makes many common sense ideas wrong including the idea that experimenters should be uncertain about the outcome of their experiments even if they know the initial conditions and equations of motion of a quantum system. However, common sense ideas change over time as people come to believe different explanations. For example, at one time people regarded the existence of witches as common sense because witches played a role in their explanations of their misfortunes and now they do not. So what did uncertainty explain in classical decision theory? In classical physics only one world exists and it evolves deterministically. It is possible that historically the reason people attached such probability to uncertainty was that the only way it could make sense for an agent to refer to more than one world in classical physics and to attach weights to them was if the agent was ignorant of some relevant feature of the laws of motion or initial conditions. In the many worlds theory the agent will experience many futures if the observable measured is unsharp relative to the state. The decision theoretic probability derived above shares all of the features that are traditionally thought to hold for the decision theoretic probability in classical decision theory except uncertainty.

As I mention above the notion of quantum Darwinism I employ only allows for perfect correlations not for imperfect correlations. I do this because as soon as imperfect correlations are introduced I have to start discussing the probability of getting the right answer out of a channel and that would defeat the purpose of writing this paper. After the Born rule has been derived using the strict notion of quantum Darwinism employed above there is nothing to prevent anyone employing it to study the imperfect, probabilistic transmission of information and the degree of classicality of specific systems as in (Ollivier *et al.*, 2004 and 2005).

An anonymous referee objected to the decision theoretic view of quantum probability on the grounds that the probability does not match the relative frequency. Probabilistic statements are tested by assuming that the relative frequency of results is similar to the probability although they usually will not match exactly (Popper, 1959, Sections 65-68). So if the relative frequency does not match the probability even approximately as will happen in some worlds, then perhaps the many worlds decision theoretic approach to probability renders probabilistic statements untestable. This objection is wrong. In the decision theoretic approach the more times an experiment is repeated, the more probable worlds in which the relative frequencies of different results approximately match the probability will become. Worlds in which the relative frequencies of different results do not approximately match the probability will become less probable. So an agent will be more rational to bet that the relative frequency will approximately match the probability than to assume that the relative frequency will not match the probability. In some worlds the rewards the players will get will not reflect this fact. However, this just means that this experiment, like every other, can go wrong, which is already true for a long list of other reasons. For example, a scientist can misinterpret experimental results, he can misunderstand the physics of the measuring instrument or the measured phenomenon, the instrument can fail to perform correctly and so on. There is no reason as to why the problem of testing probabilistic statements using relative frequencies makes science unviable while all of the other problems do not. I should also note that this problem is not unique to the many worlds theory or the decision theoretic approach to that theory. In the Copenhagen Interpretation and other collapse theories the state can collapse into a world in which relative frequencies don't match probabilities as predicted by the Born rule even approximately. The corpuscles of the pilot wave theory might move into a branch where relative frequencies don't match probabilities as predicted by the Born rule even approximately. Hidden variables theories say that quantum physics is just plain false because it is only an approximation to a deeper theory and so presumably it is also possible for the world described by these theories to deviate from the Born rule. So this problem cannot have any relevance for distinguishing the decision theoretic approach to the many worlds theory from any other theory. The fact that the Born rule can be derived from the many worlds theory using decision theory seems to indicate that it is substantially deeper than the other theories available.

**Acknowledgements**

I am indebted to David Deutsch for comments on an early draft of this paper and to anonymous referees for their helpful criticisms.